\documentclass[aps,prresearch,reprint,longbibliography,floatfix]{revtex4-2}

\usepackage[T1]{fontenc}
\usepackage[utf8]{inputenc}
\usepackage{graphicx}
\usepackage{amsmath}
\usepackage{amssymb}
\usepackage{physics}
\usepackage{siunitx}
\usepackage{hyperref}
\usepackage{color}
\usepackage{multirow}
\usepackage{booktabs}

\begin{document}

\title{Spectral Control of a Cavity-Based X-ray Free-Electron Laser via Active Mode Locking}

\author{Nanshun Huang$^1$}
\affiliation{Shanghai Advanced Research Institute, Chinese Academy of Sciences, Shanghai, China}
\author{Hanxiang Yang$^1$}
\affiliation{Shanghai Advanced Research Institute, Chinese Academy of Sciences, Shanghai, China}



\author{Haixiao Deng$^1$}
\email{denghx@sari.ac.cn}
\affiliation{Shanghai Advanced Research Institute, Chinese Academy of Sciences, Shanghai, China}

\date{\today}

\begin{abstract}
Precise spectral control in the hard X-ray regime remains a long-standing challenge that limits applications in atomic-scale science and ultrafast spectroscopy. We present an actively mode-locked cavity-based X-ray free-electron laser that achieves deterministic spectral programmability with phase-locked pulse trains and comb-like spectra, by coherently modulating the electron-beam energy. Three-dimensional time-dependent simulations predict \SI{700}{\micro\joule} total energy, \SI{30}{\giga\watt} peak power, and frequency-comb spacing of \SI{1.55}{\electronvolt} set by the modulation frequency. We further develop selective single-line amplification via undulator tapering and absolute frequency positioning through modulation-laser tuning with better than $2 \times 10^{-5}$ relative precision. Importantly, stable mode-locked operation persists under >80\% peak-to-peak cavity-reflectivity variations, substantially relaxing requirements on X-ray optics. These results establish active mode locking as a practical route to fully coherent, spectrally agile hard X-ray sources and enable new opportunities in time-resolved core-level spectroscopy, X-ray quantum optics, and precision metrology.
\end{abstract}


\maketitle

\section{Introduction}
\label{sec:intro}

The ability to coherently manipulate and shape photon spectra underpins applications spanning from optical atomic clocks and precision frequency metrology to dual-comb spectroscopy across the terahertz, infrared, and ultraviolet domains~\cite{Hansch2006,Udem2002,Diddams2020,Coddington2016}. In particular, mode-locked frequency combs and their time-programmable variants have opened new frontiers in precision metrology and quantum-limited sensing, enabling measurements that approach fundamental quantum limits~\cite{Maroju2020,Eramo2024,Caldwell2022}. Extending such coherent spectral control to hard X-rays would unlock time-resolved X-ray spectroscopy~\cite{Corkum2007,Krausz2009}, X-ray quantum optics with nuclear resonances~\cite{Palffy2009,Heeg2015,Gunst2016,Haber2016}, and attosecond X-ray pulse generation~\cite{Zholents2004,Robles2025}, while strengthening tests of fundamental physics through precision measurements~\cite{Huang2021,Safronova2018}.

Modern X-ray free-electron lasers (XFELs) based on self-amplified spontaneous emission (SASE)~\cite{Pellegrini2016,Emma2010,Ishikawa2012,Kang2017,swissFEL2020,Decking2020} deliver very high peak brightness but lack full temporal coherence, which makes accurate spectral control difficult. To improve coherence, external seeding~\cite{Stupakov2009,Xiang2009,FERMIFEL2012,SXFEL_2022_EEHC} and self-seeding schemes~\cite{Geloni2011,Amann2012,Inoue2019} have been widely developed and used to narrow the bandwidth and enhance longitudinal coherence. Building on these developments, several spectral-control techniques—including mode-locked FEL pulses~\cite{Zholents2004,Saldin2006,Thompson2008}—have been demonstrated in the soft X-ray~\cite{Gauthier2016} and hard X-ray ranges~\cite{Hu2025}. However, these methods typically operate in a single pass and are challenging to implement at high repetition rates. In addition, they are highly sensitive to electron jitter in beam energy and arrival time, which restricts the achievable precision and stability of spectral control.

The cavity-based X-ray free-electron laser (CBXFEL) offers a compelling platform for fully coherent X-rays and advanced spectral control~\cite{Kim2008,XFELO_harmonic_2012_deng}. Unlike single-pass XFELs, CBXFELs recirculate and amplify X-ray pulses in a high-quality optical cavity over many passes, inherently providing full transverse coherence, narrow bandwidth, and enhanced spectral stability. Two main configurations exist: X-ray FEL oscillators, which operate in a low-gain regime and require many round trips to reach saturation, and X-ray regenerative amplifier FELs (XRAFELs)~\cite{Huang2006,Freund2019,Marcus2020}, which utilize high gain to achieve saturation with fewer passes while enabling stronger outcoupling. Recent developments have brought CBXFEL to realization~\cite{Shvydko2010,Stoupin2010,Rauer2023}. At the European XFEL, the successful demonstration of first multi-pass lasing in a hard X-ray cavity marks a major milestone, establishing the feasibility of cavity-based XFELs~\cite{Rauer2025}. However, a practical mechanism for active spectral control with the required precision and agility has remained elusive.

\begin{figure*}[t]
\centering
\includegraphics[width=1.8\columnwidth]{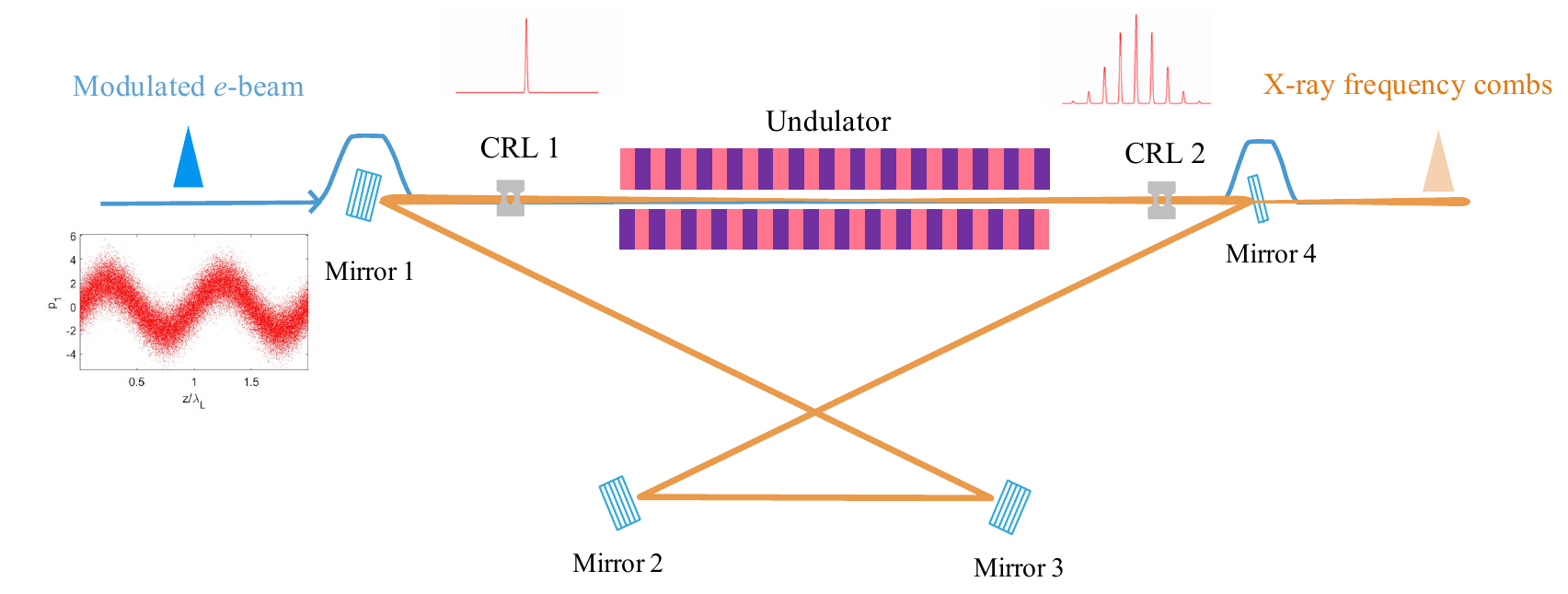}
\caption{Concept of actively mode-locked CBXFEL. An external infrared laser (\SI{800}{\nano\meter}) imposes periodic energy modulation in an electron. The modulated electron beam interacts with the intracavity monochromatic X-ray field in the main undulator, producing a phase-locked pulse train (frequency comb). }
\label{fig:scheme}
\end{figure*}

To address this limitation, we propose a robust method for active spectral control in hard X-rays based on coherent electron-beam energy modulation in a CBXFEL. By integrating energy modulation with cavity-based amplification, the scheme establishes active mode locking and produces controllable comb-like spectra. The underlying physics is analogous to active mode locking in optical lasers~\cite{Mocker1965,DeMaria1966,Haus2000}, where sinusoidal amplitude modulation at the cavity-mode spacing produces phase locking of longitudinal modes. Here, modulation of the lasing medium—the relativistic electron beam—imprints a corresponding modulation on the intracavity field at the laser frequency, leading to well-defined comb sidebands. This approach exploits the monochromatic feature of the X-ray crystal cavity while simplifying implementation by eliminating chicanes that repeatedly shift the radiation phase in previously proposed schemes~\cite{Thompson2008}. We demonstrate feasibility through numerical simulations using realistic parameters accessible at facilities such as SHINE~\cite{Zhao2017,Li2018}.

The paper is organized as follows: Section~\ref{sec:concept} presents the active mode-locking mechanism; Section~\ref{sec:results} reports FEL simulations of comb formation and output characteristics; Section~\ref{sec:control} demonstrates advanced spectral control capabilities; Section~\ref{sec:robustness} analyzes cavity tolerance and practical implementation considerations; and we conclude with a discussion of scientific impact and future prospects.

\section{Active Mode-Locking Mechanism}
\label{sec:concept}

Our approach establishes active mode-locking in hard X-rays by synergistically combining coherent electron-beam energy modulation with cavity-based amplification. The mechanism operates through three fundamental processes: (1) periodic energy modulation of relativistic electrons by an external laser, (2) the modulated electrons amplify the monochromatic X-ray radiation, generating a phase-locked pulse train (frequency comb), and (3) the crystal cavity selects specific frequency components through intracavity feedback, creating a self-reinforcing mode-locking cycle.

This mechanism can be analytically understood using FEL theory~\cite{Pellegrini2016,kim_synchrotron_2017}. In the small-signal regime, the coherent energy modulation imparted to an electron after traversing the modulator undulator is given by

\begin{equation}
\Delta \gamma = M \sin(\omega_{L} t + \phi_{0}),
\label{eq:energy_mod}
\end{equation}
where $M$ denotes the normalized modulation amplitude, $\omega_{L}$ is the laser angular frequency, and $\phi_{0}$ is the initial laser phase. This sinusoidal modulation creates a periodic variation in the electron energy that directly translates to modulated gain in the main undulator.

The evolution of the intracavity radiation field over successive round trips must account for mirror reflectivity, accumulated phase shifts, and amplification from the modulated electron beam. The field amplitude after the $m$-th round trip is related to that of the previous pass by~\cite{kim_synchrotron_2017}
\begin{equation}
A_{m}(t) = A_{m-1}(t)\, e^{i\phi_{rt}} \sqrt{R}\, G_{m}(t),
\label{eq:cavity_evolution}
\end{equation}
where $R$ is the mirror reflectivity, $\phi_{rt}$ is the round-trip phase shift, and $G_{m}(t)$ denotes the single-pass gain factor, which depends on the instantaneous beam energy. A convenient representation for the gain is $G_m(t) \approx g_0 \exp[\Delta \gamma (t)]$, where $g_0$ is the baseline gain~\cite{Pellegrini2016}.

Substituting Eq.~\eqref{eq:energy_mod} into the gain function and applying the Jacobi--Anger expansion,
\[
e^{i M \sin(\omega_{L} t + \phi_{0})} = \sum_{q=-\infty}^{\infty} J_{q}(M)\, e^{i q (\omega_{L} t + \phi_{0})},
\]
the intracavity radiation field can be expressed as
\begin{align}
A(t) &\propto e^{i\omega_{0} t}
        \sum_{q=-\infty}^{\infty} C_{q}\,
        e^{i q \omega_{L} t}, \\
A(\omega) &\propto \sum_{q=-\infty}^{\infty} C_{q}\,
        \delta \Bigl(\omega - \bigl(\omega_{0}+q\omega_{L}\bigr)\Bigr),
        \label{eq:A_time}
\end{align}
where $C_{q} = J_{q}(M)\, e^{i q \phi_{0}}$, $J_q$ is the $q$-th order Bessel function of the first kind, and $\omega_{0}$ is the central cavity frequency. The series in Eq.~(\ref{eq:A_time}) represents a periodic function with period $T_{L}=2\pi/\omega_{L}$. In the frequency domain, this corresponds to discrete spectral lines at frequencies $\omega = \omega_0 + q \omega_L$, forming a well-defined frequency comb.

Consequently, the sinusoidal energy modulation of the electron beam imposes periodic gain modulation on the radiation field, acting as an active mode-locking mechanism. This forces the field $A(t)$ to evolve into a stable mode-locked state characterized by a frequency comb spectrum with precisely controlled tooth spacing. In this configuration, the output radiation is not a continuous pulse but rather a phase-locked pulse train comprising a superposition of discrete frequency modes, with the electron-beam-induced modulation serving as the active driver that maintains coherent phase relationships among all spectral components.

The system provides two complementary approaches for precise spectral control:

\textit{Selective spectral line amplification}: Undulator tapering introduces frequency-dependent gain that selectively amplifies individual spectral lines. A controlled decrease in the undulator parameter $K(z)$ along the undulator length tilts the gain profile, enabling single-line operation without compromising overall system stability.

\textit{Absolute frequency positioning}: The modulation laser frequency directly determines both comb tooth spacing and absolute spectral positions, while crystal Bragg mirrors ensure precise frequency alignment. With stabilized laser systems, this approach enables meV-level precision in X-ray frequency control—a capability essential for precision spectroscopy applications requiring both high resolution and absolute frequency accuracy.

\section{Mode-locking Results and Performance Characteristics}
\label{sec:results}

We demonstrate the feasibility of actively mode-locked hard X-ray frequency combs through three-dimensional time-dependent simulations using GENESIS~\cite{Reiche1999} coupled with OPC for custom cavity propagation modeling~\cite{XFELO_OPC_2006} and BRIGHT~\cite{BRIGHT_huang_2019} for Bragg mirror simulation. Our actively mode-locked CBXFEL employs realistic parameters achievable with current technology, such as those at SHINE, supporting the practical implementation of this approach.

The superconducting linear accelerator delivers electron bunches with \SI{8}{\giga\electronvolt} energy, \SI{100}{\pico\coulomb} charge, normalized transverse emittance of \SI{0.4}{\micro\meter}, relative energy spread $\sigma_\gamma/\gamma_0 = 1\times10^{-4}$, and \SI{1}{\mega\hertz} repetition rate. This configuration produces optimal energy modulation (dimensionless amplitude $M \approx 5$) corresponding to 0.05\% of the beam energy while preserving the beam quality essential for efficient FEL amplification. This modulation with an external seed can be readily achieved at high repetition rates~\cite{Yan2021}. The high-quality X-ray cavity features a \SI{300}{\meter} round-trip length employing diamond crystal Bragg mirrors with reflectivity $R = 0.99$ at \SI{12.9}{\kilo\electronvolt}, while the \SI{120}{\meter} main undulator section provides amplification of the intracavity field.

\begin{figure}[t]
\centering
\includegraphics[width=\columnwidth]{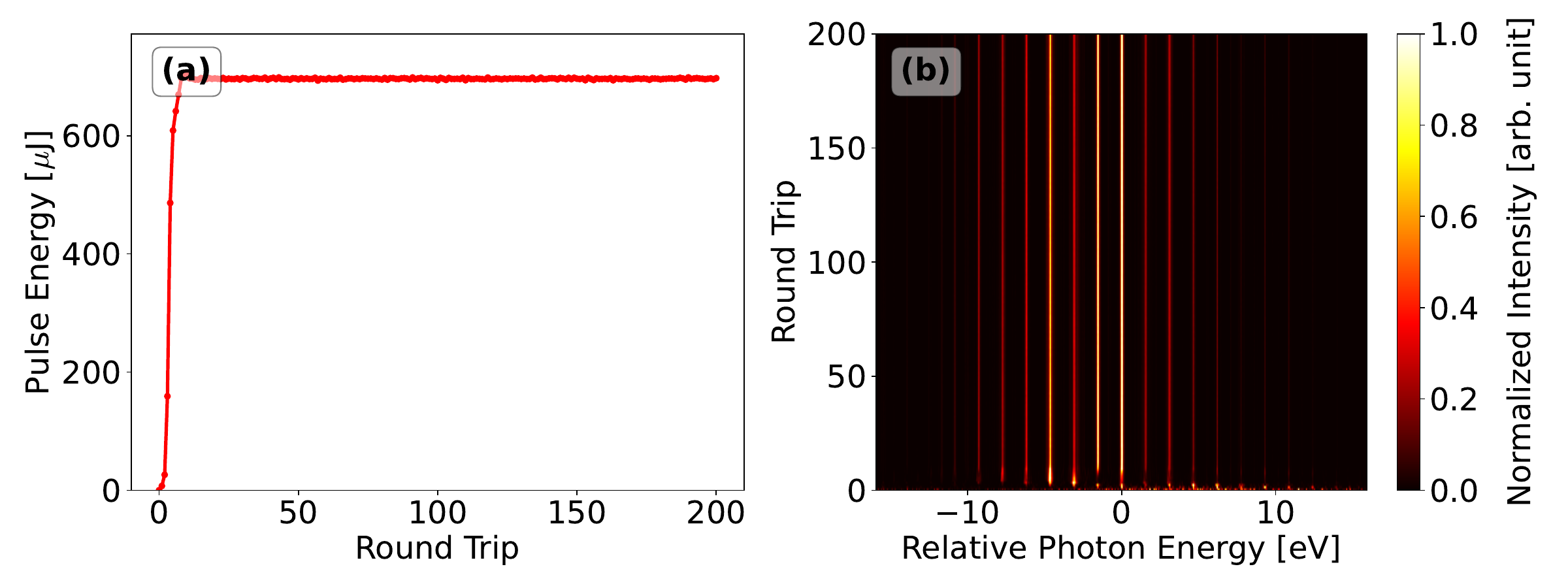}
\caption{Evolution of actively mode-locked CBXFEL operation. (a) Pulse energy buildup over 200 round trips (\SI{0.2}{\milli\second}), demonstrating rapid convergence to steady-state operation at \SI{700}{\micro\joule}. (b) Spectral evolution showing the emergence of a well-defined frequency-comb structure with tooth spacing of \SI{1.55}{\electronvolt} corresponding to the \SI{800}{\nano\meter} modulation laser wavelength.}
\label{fig:comb_evo}
\end{figure}

\begin{figure}[t]
\centering
\includegraphics[width=\columnwidth]{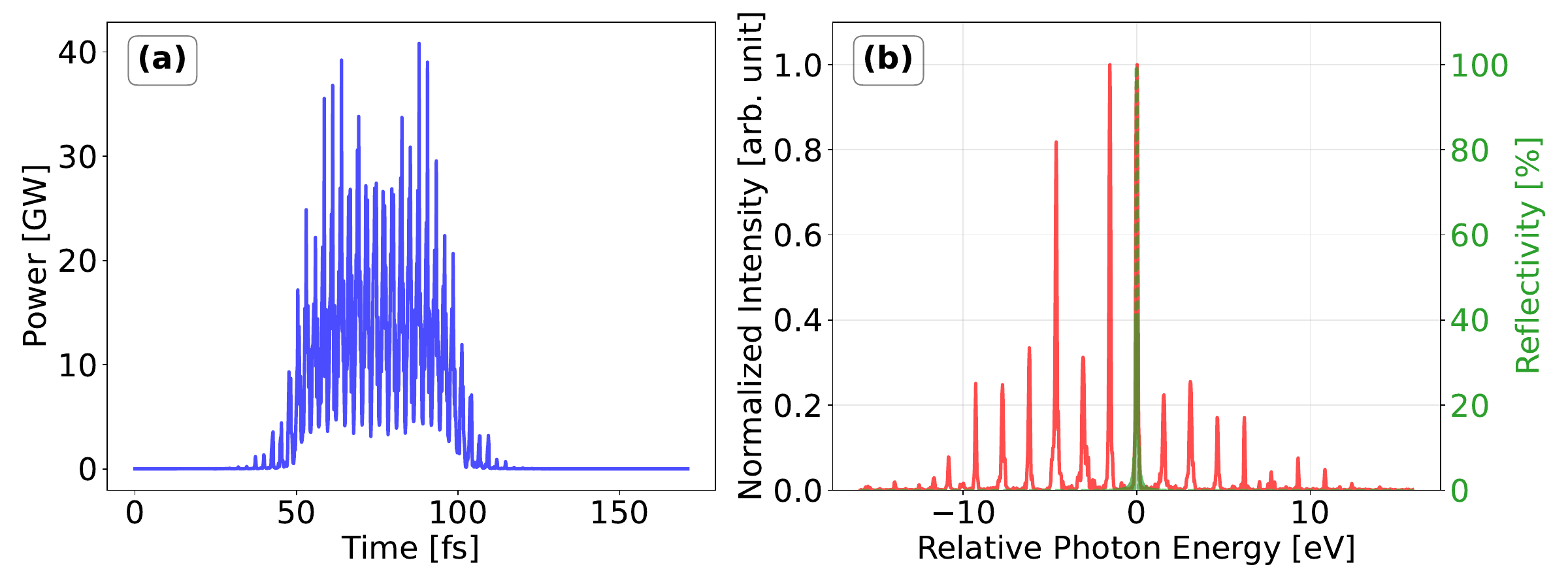}
\caption{Steady-state characteristics of the actively mode-locked CBXFEL. (a) Temporal power profile of the X-ray output showing a train of femtosecond pulses ($\sim$\SI{1.5}{\femto\second} FWHM) with peak power exceeding \SI{30}{\giga\watt}. (b) Corresponding frequency-domain spectrum exhibiting a well-defined comb structure with normalized intensity (red line) and cavity mirror reflectivity profile (green line).}
\label{fig:pulse_p}
\end{figure}

We first demonstrate the mode-locked output, producing a frequency comb-like spectral output. The performance of the proposed mode-locked CBXFEL is illustrated in Fig.~\ref{fig:comb_evo} and Fig.~\ref{fig:pulse_p}. 

Fig.~\ref{fig:comb_evo} demonstrates the successful initiation and evolution of actively mode-locked operation in the cavity-based X-ray free-electron laser. The energy buildup dynamics shown in Fig.~\ref{fig:comb_evo}(a) reveal rapid convergence from zero initial conditions to a steady-state pulse energy of \SI{700}{\micro\joule} over 200 round trips, corresponding to approximately \SI{0.2}{\milli\second} of operation. The transition occurs within the first ten round trips, indicating robust mode-locking dynamics and efficient energy extraction from the gain medium.

Simultaneously, Fig.~\ref{fig:comb_evo}(b) presents the spectral evolution throughout the mode-locking process as a two-dimensional intensity map. The emergence of a well-defined frequency-comb structure is clearly visible, with equally spaced spectral lines appearing as the system reaches steady state. The comb tooth spacing of \SI{1.55}{\electronvolt} precisely corresponds to the \SI{800}{\nano\meter} wavelength of the modulation laser, while the spectral lines extend over approximately \SI{20}{\electronvolt} bandwidth, demonstrating broadband coherent X-ray generation with exceptional spectral control.


Fig.~\ref{fig:pulse_p} characterizes the steady-state temporal and spectral performance of the actively mode-locked CBXFEL system. The temporal power profile in panel (a) reveals a train of ultrashort pulses with exceptional \SI{1.5}{\femto\second} FWHM duration and extraordinary peak power exceeding \SI{30}{\giga\watt}. By adjusting the energy modulation amplitude, the single pulse duration can be further compressed to the attosecond regime with reduced pulse energy.

The corresponding frequency-domain spectrum in Fig.~\ref{fig:pulse_p}(b) exhibits a well-defined comb structure with multiple discrete peaks, confirming stable phase-locking between longitudinal modes. The normalized intensity distribution (red line) shows excellent agreement with the cavity mirror reflectivity profile (green line), with the spectral envelope spanning approximately \SI{1.55}{\electronvolt}. The relative intensities of the comb teeth are naturally modulated by the cavity mirror reflectivity profile, which peaks near the central photon energy at \SI{12.9}{\kilo\electronvolt}. These comb lines can also be controlled by parameters such as the crystal reflectivity setting, the overall cavity efficiency, and the modulation depth.


\section{Advanced Spectral Control}
\label{sec:control}

Beyond generating stable frequency combs, our actively mode-locked CBXFEL provides unprecedented control over X-ray spectral properties through two synergistic mechanisms: selective single-line amplification via undulator tapering and precise absolute frequency control through laser stabilization. These complementary approaches enable tailored spectral output for diverse scientific applications while maintaining the exceptional temporal characteristics of mode-locked operation.

\subsection{{Selective spectral line amplification}}

While the fundamental mode-locking mechanism produces a multi-line frequency comb spanning multiple electronvolts, many precision spectroscopy applications require single-frequency operation with narrow linewidth and high spectral purity. We achieve this transformation through controlled undulator tapering~\cite{Fawley2002,Orzechowski1986,Schneidmiller2015,Mak2015}, which introduces frequency-dependent gain that selectively amplifies a chosen spectral line while systematically suppressing adjacent comb teeth.

The undulator parameter $K(z)$ is gradually reduced along the undulator length. This controlled variation creates a frequency-dependent gain profile that peaks at a specific photon energy determined by the tapering strength, effectively filtering the frequency comb through gain discrimination rather than passive spectral filtering. Furthermore, since it is unnecessary to excite such a large number of comb lines, the energy modulation amplitude can be reduced, which in turn further enhances the pulse energy.

\begin{figure}[t]
\centering
\includegraphics[width=0.95\columnwidth]{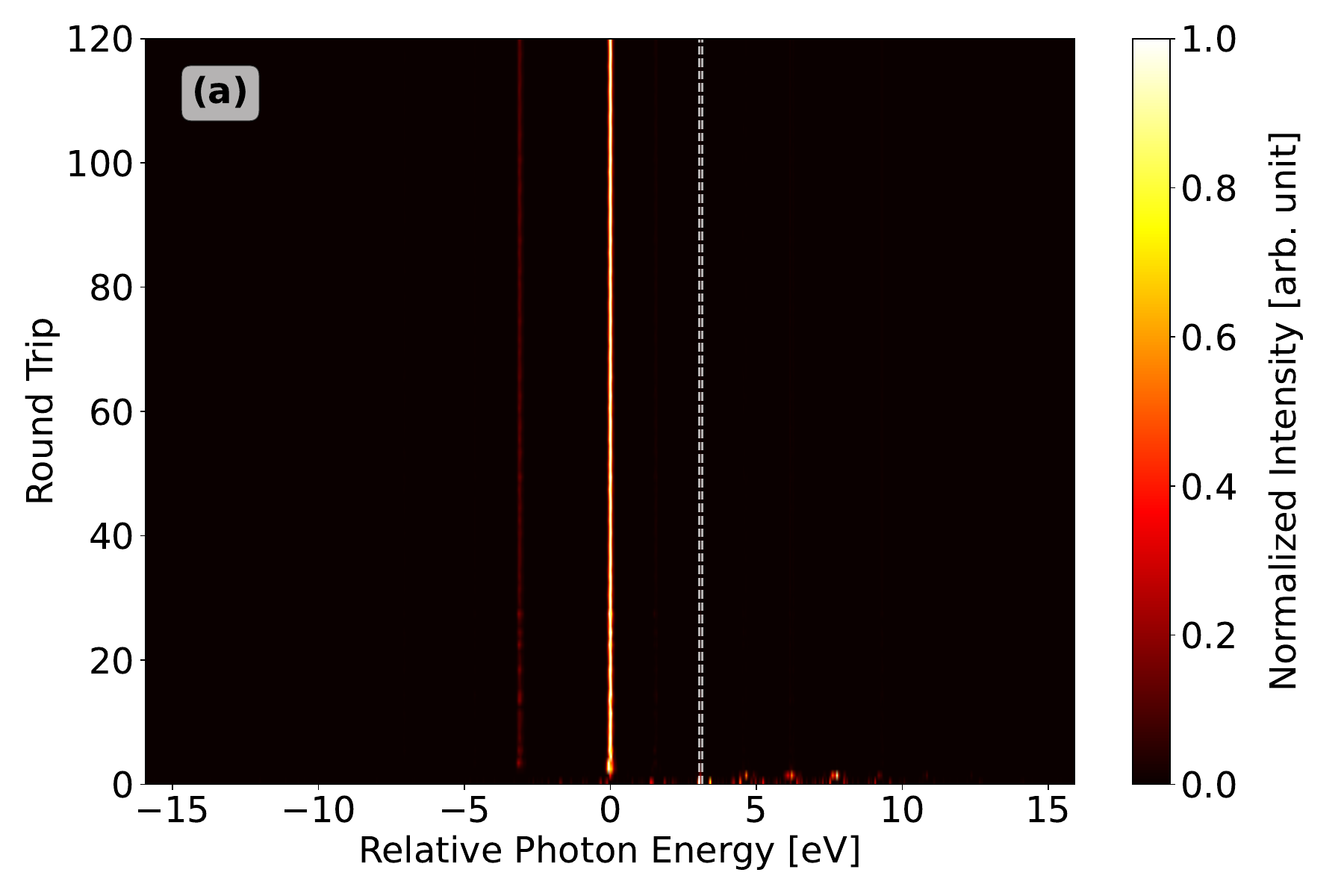}
\includegraphics[width=0.95\columnwidth]{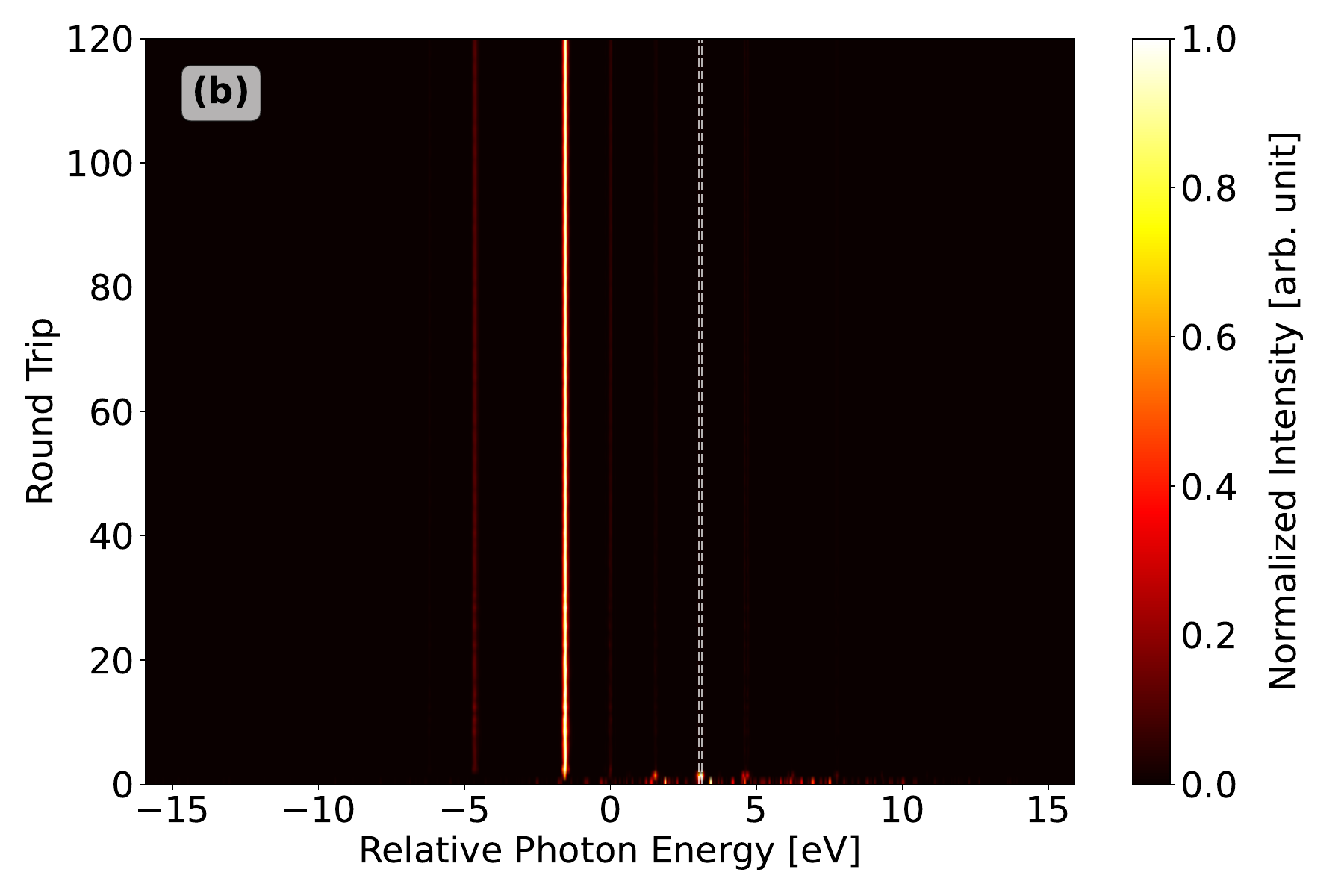}
\caption{Selective single-line amplification via undulator tapering. (a) Amplification of the central comb tooth, achieving >95\% power concentration. (b) Selection of off-center teeth, demonstrating versatile spectral control across the comb bandwidth. The white dashed line marks the Bragg reflective window of the cavity is applied.}
\label{fig:single_line_control}
\end{figure}

Fig.~\ref{fig:single_line_control} demonstrates the spectral control capability achieved through undulator tapering in the actively mode-locked CBXFEL system. The two panels illustrate the selective amplification of individual comb teeth with exceptional spectral purity and power concentration. Panel (a) shows the evolution of the spectral profile when the undulator taper is optimized for amplifying the central comb tooth at the center photon energy. The initially broad comb spectrum, visible in the early round trips, rapidly converges to a single dominant spectral line. At steady state, the system achieves greater than 95\% power concentration in the selected tooth, with neighboring teeth suppressed. The white dashed line marks the Bragg reflective window of the cavity. The resulting narrow-bandwidth output maintains the high peak power characteristics of the mode-locked operation while providing exceptional spectral brightness.

Panel (b) illustrates the versatility of this spectral control technique by demonstrating the selection of an off-center comb tooth at approximately \SI{-1.5}{\electronvolt} relative photon energy. The evolution follows a similar pattern, with the system transitioning from the initial multi-tooth comb structure to single-line dominance. The successful amplification of off-center teeth confirms that the tapering mechanism can be tuned across the entire comb bandwidth, enabling flexible wavelength selection without compromising the stability of the mode-locked operation. 

This technique transforms the broadband comb into a tunable, quasi-monochromatic X-ray source while preserving the temporal coherence and high peak power inherent to the mode-locked operation. The smooth evolution from multi-tooth to single-line operation, without any signs of instability or mode competition, validates the robustness of this approach for precision spectroscopy applications requiring both high spectral resolution and high photon flux. This method maintains robust mode-locking operation while providing precise spectral shaping.

\subsection{Absolute frequency positioning}

The second control mechanism provides meV-level precision in absolute frequency positioning through sophisticated stabilization of both the modulation laser and the cavity length. Since the comb tooth positions are directly determined by the modulation laser frequency ($f_{\text{tooth}} = f_0 + n \cdot f_{\text{mod}}$), precise laser stabilization combined with cavity length control enables unprecedented frequency stability in the hard X-ray regime. This dual-stabilization approach creates a metrological chain that transfers optical frequency standards to the X-ray domain.

\begin{figure}[t]
\centering
\includegraphics[width=0.95\columnwidth]{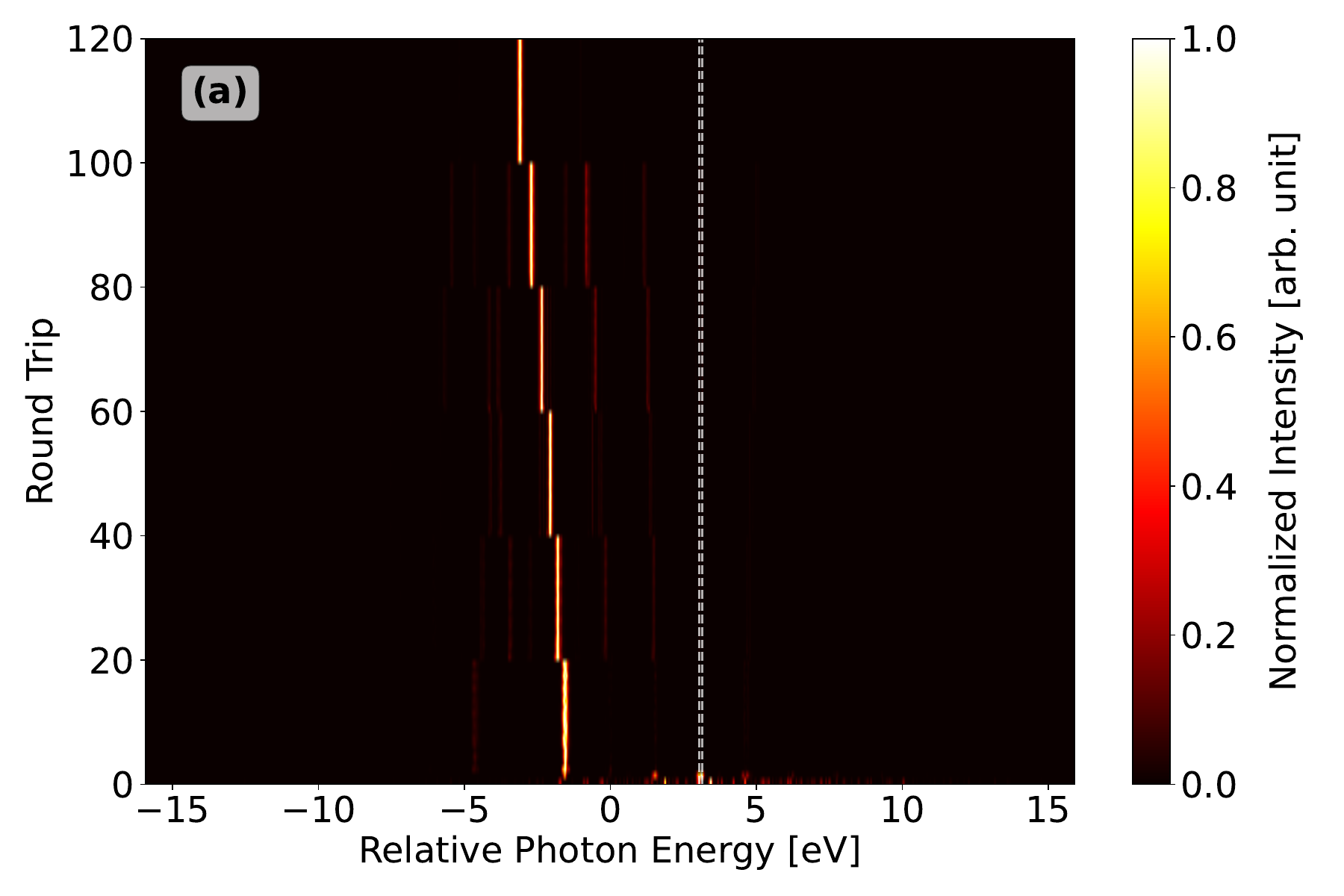}
\includegraphics[width=0.95\columnwidth]{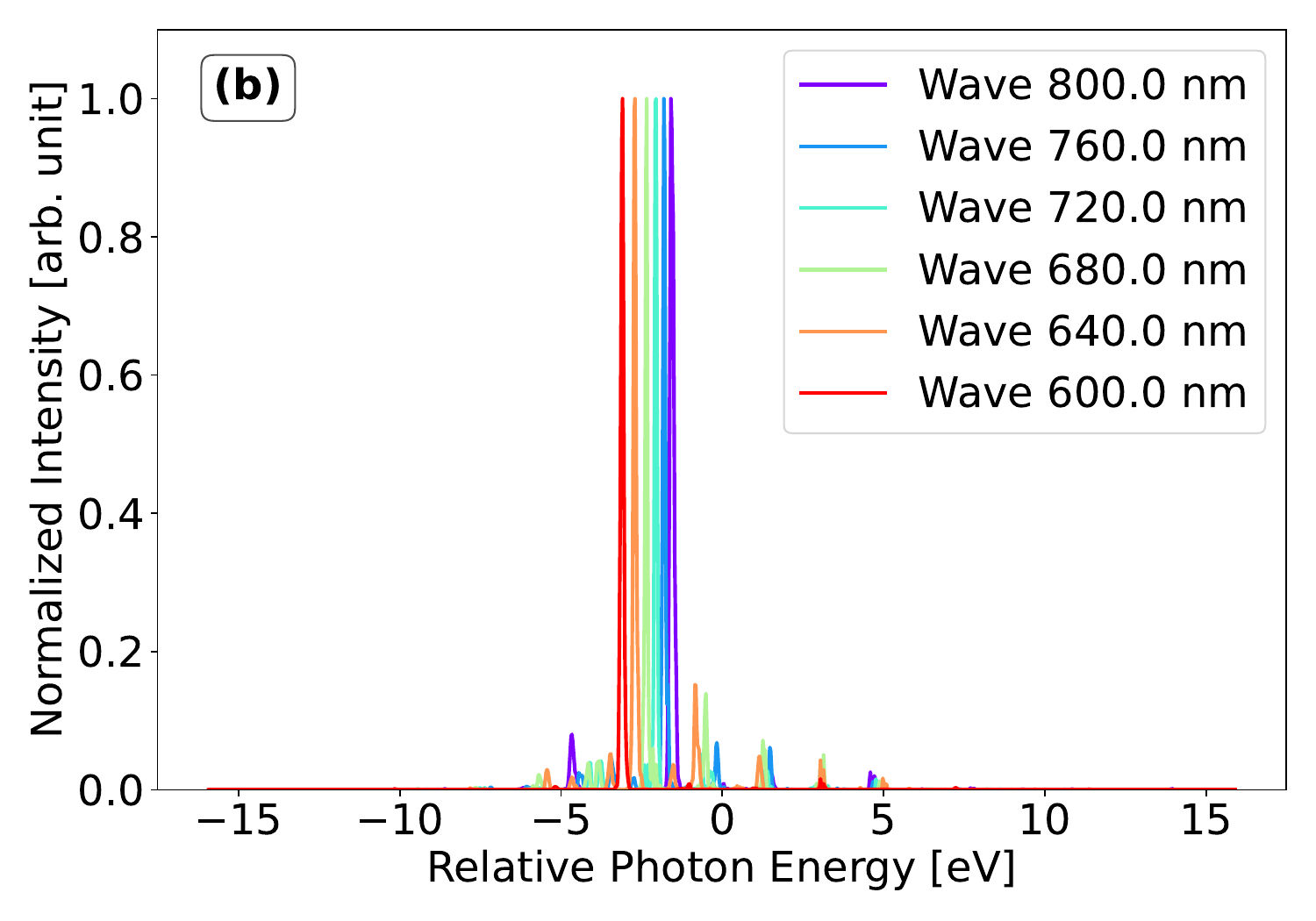}
\caption{Demonstration of precision frequency control. (a) Spectral evolution showing the dynamic response to modulation wavelength tuning, with the white dashed line indicating the point of wavelength adjustment. (b) Overlaid spectra at different modulation laser wavelengths from \SI{600}{\nano\meter} to \SI{800}{\nano\meter}, demonstrating continuous tunability of the selected comb tooth at \SI{12.9}{\kilo\electronvolt}—a capability essential for precision X-ray spectroscopy. The white dashed line marks the Bragg reflective window of the cavity.}
\label{fig:line_tune}
\end{figure}

Fig.~\ref{fig:line_tune} showcases the exceptional frequency control capabilities of the actively mode-locked CBXFEL system through modulation wavelength tuning. This demonstration reveals the ability to provide continuous, high-precision spectral adjustment crucial for advanced X-ray spectroscopy and metrological applications.

Fig.~\ref{fig:line_tune}(a) presents the temporal evolution of the spectral profile as the modulation laser wavelength is dynamically adjusted. After every \SI{20}{} round trips, the modulation wavelength is changed by \SI{40}{\nano\meter}, triggering a smooth transition to a new spectral position. The system rapidly re-establishes single-line operation at the new frequency with approximately \SI{1}{} round trip, demonstrating both the responsiveness and stability of the frequency control mechanism. The continuous spectral shift without loss of mode-locking quality or power concentration validates the robustness of this tuning approach.

Fig.~\ref{fig:line_tune}(b) provides a comprehensive view of the tuning range by overlaying the steady-state spectra obtained at six different modulation wavelengths spanning from \SI{600}{\nano\meter} to \SI{800}{\nano\meter}. Each colored trace represents a fully optimized single-line output at a specific modulation wavelength. The systematic shift in peak position with changing modulation wavelength demonstrates precise and predictable frequency control. The tooth positions shift continuously across an approximately \SI{1.6}{\electronvolt} range with $n=3$ while maintaining consistent peak intensity and spectral purity.

The remarkable aspect of this demonstration is the achievement of meV-level precision at the operating photon energy of \SI{12.9}{\kilo\electronvolt}. This represents a relative precision of better than $2 \times 10^{-5}$. Through more precise laser wavelength tuning, we have the potential to achieve a tuning precision of $10^{-6}$, which significantly improves tuning precision compared with traditional methods and establishes a new paradigm for X-ray metrology. The narrow linewidth of each peak (less than \SI{100}{\milli\electronvolt} FWHM) and the ability to position these peaks with meV accuracy enable resolution of fine spectroscopic features that would be inaccessible with conventional X-ray sources.

This exceptional precision opens transformative opportunities across multiple scientific domains. In precision atomic physics, the ability to tune X-ray frequencies with meV accuracy enables direct probes of core electronic transitions. For time-resolved spectroscopy, the combination of femtosecond temporal resolution with meV spectral precision allows unprecedented studies of ultrafast dynamics in complex materials, including chemical shifts, charge transfer processes, and coherent phonon dynamics. The precise tunability, combined with the high peak power and temporal coherence of the mode-locked pulses, positions this system as an ideal tool for next-generation X-ray quantum optics experiments~\cite{Palffy2009,Heeg2015,Gunst2016,Haber2016}, including X-ray photon entanglement and quantum state manipulation at atomic length scales.

\section{Robustness and Practical Implementation}
\label{sec:robustness}

A critical advantage of our actively mode-locked CBXFEL scheme lies in its exceptional robustness against cavity imperfections—a transformative feature that fundamentally relaxes the stringent requirements that have limited CBXFEL development~\cite{Huang2020}. In the first lasing experiment of the cavity-based XFEL at European XFEL~\cite{Rauer2025}, significant performance degradation was observed due to thermal loading. Unlike passive cavity-based approaches that demand near-perfect optical components with extraordinary stability, our active mode-locking mechanism maintains stable operation even under severe cavity efficiency variations that would destabilize conventional systems.

\begin{figure}[t]
\centering
\includegraphics[width=\columnwidth]{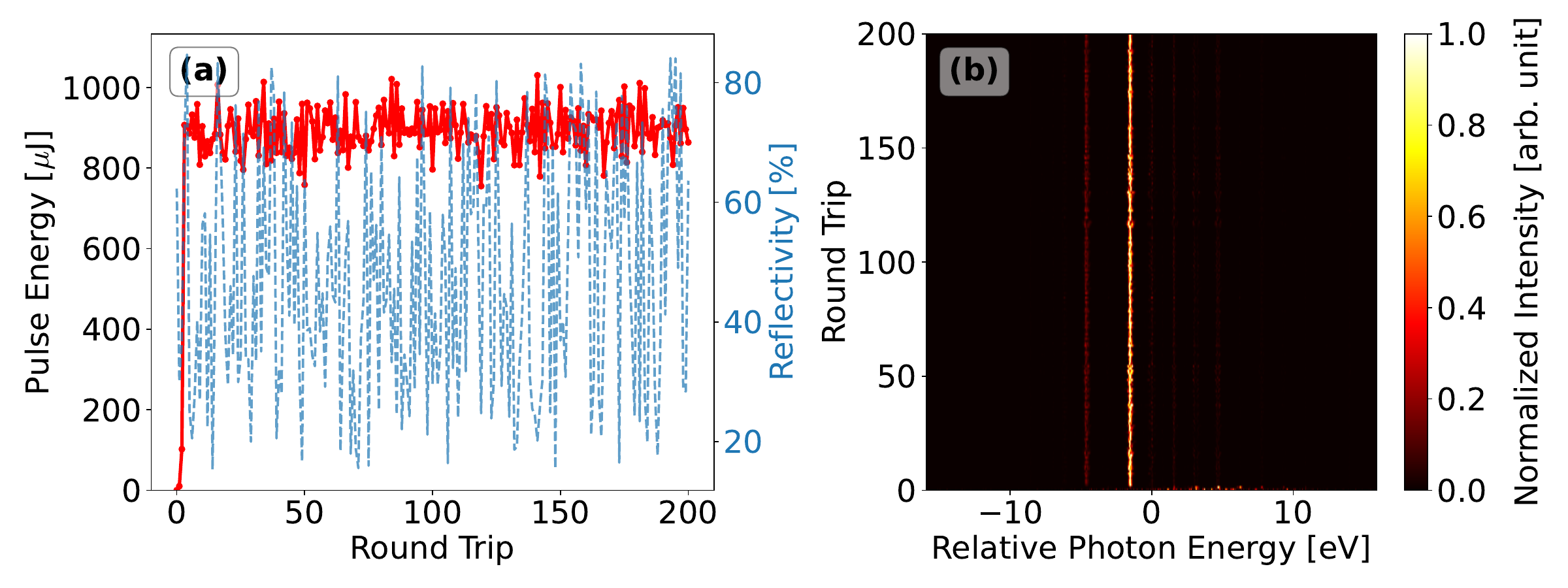}
\caption{Tolerance to cavity reflectivity fluctuations. (a) Pulse energy (red) remains at \SI{800}{\micro\joule} despite >80\% peak-to-peak fluctuations in cavity efficiency (blue dashed). (b) The spectrum maintains a narrow single line throughout, confirming robust mode-locked operation under severe cavity variations.}
\label{fig:single1_jitter}
\end{figure}

Fig.~\ref{fig:single1_jitter} presents a demonstration of robustness against cavity mirror imperfections, addressing one of the most critical challenges for practical implementation of CBXFEL. Fig.~\ref{fig:single1_jitter}(a) displays the temporal evolution of both pulse energy (red solid line) and cavity mirror reflectivity (blue dashed line) over \SI{200}{} round trips. The reflectivity profile exhibits extreme fluctuations, varying from approximately 10\% to over 90\%, representing peak-to-peak variations exceeding 80\% of the nominal value—fluctuations far beyond what conventional cavity systems could tolerate. These severe perturbations simulate realistic operational conditions including mirror surface irregularities, thermal distortions, and alignment imperfections that are inevitable in high-power X-ray optics environments. Remarkably, despite these dramatic perturbations, the pulse energy maintains extraordinary stability around \SI{800}{\micro\joule} with only modest fluctuations. The energy variations remain within $\pm$10\% of the mean value, demonstrating unprecedented immunity to cavity losses and establishing a new paradigm for robust X-ray cavity operation.

Fig.~\ref{fig:single1_jitter}(b) shows the corresponding spectral evolution throughout the same challenging period. The spectrum maintains a clean, single-line profile centered at \SI{-1.5}{\electronvolt} relative photon energy, with consistent intensity and minimal spectral broadening throughout the entire sequence. The complete absence of mode competition, or unwanted sidebands—even during periods of extreme reflectivity variation—confirms that the fundamental mode-locking mechanism remains intact and stable. 

The underlying stability mechanism can be attributed to the synergistic interplay between gain saturation dynamics and the active mode-locking process, which together provide a sophisticated self-regulating feedback system. When reflectivity drops dramatically, reducing intracavity power, the unsaturated gain automatically increases to compensate for the losses. Conversely, periods of high reflectivity lead to stronger gain saturation, preventing runaway power growth and maintaining equilibrium. This natural compensation mechanism, enhanced by the phase-locking stability provided by active modulation, creates a robust operational equilibrium that tolerates significant cavity imperfections while preserving output quality.

This demonstration has profound implications for practical CBXFEL implementation and widespread adoption. The demonstrated tolerance to reflectivity variations exceeding 80\% suggests that actively mode-locked CBXFELs can operate reliably with currently achievable mirror technology, eliminating the need for prohibitively expensive perfect optics. Furthermore, this exceptional robustness significantly extends the operational lifetime of cavity optics, as gradual degradation will not catastrophically impact performance or require frequent replacement. 

\section{Conclusions}
\label{sec:conclusion}

We introduced an actively mode-locked cavity-based X-ray free-electron laser that provides programmable spectral control at \SI{12.9}{\kilo\electronvolt} by combining coherent electron beam energy modulation with cavity feedback. Three-dimensional time-dependent simulations predict \SI{30}{\giga\watt} peak power, \SI{700}{\micro\joule} pulse energy, and a frequency comb spacing of \SI{1.55}{\electronvolt} set by the modulation laser frequency. Spectral shaping is enabled by two complementary mechanisms: single-line amplification via undulator tapering, which concentrates >95\% of the power into a selected comb tooth while maintaining temporal coherence, and absolute frequency tuning via modulation laser tuning, which sets the absolute frequency with better than $2 \times 10^{-5}$ relative precision at hard X-ray energies. Mode-locked operation remains stable under peak-to-peak cavity reflectivity variations exceeding 80\%, substantially relaxing requirements on X-ray optics and improving practical feasibility. These results establish active mode locking as a practical route to fully coherent, spectrally programmable hard X-ray sources for time-resolved core-level spectroscopy, X-ray quantum optics, and precision metrology.

\section*{Acknowledgments}

This work was supported by the National Natural Science Foundation of China (Grant No. 12125508), the National Key Research and Development Program of China (Grant Nos. 2024YFA1612101 and 2024YFA1612104), the CAS Project for Young Scientists in Basic Research (Grant No. YSBR-042), and the Shanghai Pilot Program for Basic Research - Chinese Academy of Sciences, Shanghai Branch (Grant No. JCYJ-SHFY-2021-010). The authors acknowledge valuable discussions with colleagues at the Shanghai Advanced Research Institute and thank the SHINE project team for technical support and facility parameters.


%

\end{document}